\documentclass[twocolumn]{emulateapj}
\usepackage{natbib}
\usepackage{comment}
\usepackage{mathrsfs}

\def\ba{\begin{eqnarray}}
\def\ea{\end{eqnarray}}
\usepackage{graphicx, amsmath, amsthm, amssymb,color}

\newcommand{\asypole}{{\bf C3PO}}

\newcommand\tmag{$\mathcal{T}\ $}
\newcommand\teff{${T_{\rm eff}}\ $}

\newcommand{\figr}[1]{Figure.~\ref{fig:#1}}

\newcommand{\secr}[1]{\S\ref{sec:#1}}

\shorttitle{TESS Simulation}
\shortauthors{Huang et al}

\begin{document}

\title{Expected Yields of Planet discoveries from the TESS primary and extended missions}

\author{
Chelsea X.~Huang\altaffilmark{1,2},
Avi Shporer\altaffilmark{1},
Diana Dragomir\altaffilmark{1,3},
Michael Fausnaugh\altaffilmark{1}, 
Alan M.~Levine\altaffilmark{1}, 
Edward H. Morgan \altaffilmark{1},
Tam Nguyen\altaffilmark{4, 5},
George R. Ricker\altaffilmark{1}, 
Matt Wall\altaffilmark{1},
Deborah F. Woods\altaffilmark{6},
Roland K. Vanderspek\altaffilmark{1}
}

\altaffiltext{1}{Department of Physics, and Kavli Institute for Astrophysics and Space Research, Massachusetts Institute of Technology, Cambridge, MA 02139, USA}
\altaffiltext{2}{Juan Carlos Torres Fellow}
\altaffiltext{3}{NASA Hubble Fellow}
\altaffiltext{4}{Department of Aeronautics and Astronautics, Massachusetts Institute of Technology}
\altaffiltext{5}{NSF Fellow}
\altaffiltext{6}{MIT Lincoln Laboratory}

\begin{abstract}
    We present a prediction of the transiting exoplanet yield of the TESS primary mission, in order to guide follow-up observations and science projects utilizing TESS discoveries. Our new simulations differ from previous work by using (1) an updated photometric noise model that accounts for the nominal pointing jitter estimated through simulation prior to launch,  (2) improved stellar parameters based on Gaia mission Data Release 2, (3) improved empirically-based simulation of multi-planet systems, (4) a realistic method of selecting targets for 2-minute exposures, and (5) a more realistic geometric distortion model to determine the sky region that falls on TESS CCDs. We also present simulations of the planet yield for three suggested observing strategies of the TESS extended mission.
We report $\sim 10^4$ planets to be discovered by the TESS primary mission, as well as an additional $\sim 2000$ planets for each year of the three extended mission scenarios we explored. We predict that in the primary mission, TESS will discover about 3500 planets with Neptune size and smaller, half of which will orbit stars with TESS magnitudes brighter than 12. Specifically, we proposed a new extended mission scenario that centers Camera 3 on the ecliptic pole ({\bf C3PO}), which will yield more long period planets as well as moderately irradiated planets that orbit F, G, and K stars.

\end{abstract}

\section{Introduction}
\label{sec:intro}
The successful launch of the Transiting Exoplanet Survey Satellite \citep[TESS;][]{Ricker:2014, Ricker:2015} on 2018 April 18 marked the beginning of a new era for transiting exoplanet science. For the next two years, TESS will provide high precision time series photometric observations of bright stars across almost the entire sky. The TESS mission is designed to discover new planets orbiting bright stars spanning a wide range of spectral classes. 

The brightness of the host stars will enable follow-up studies that can measure detailed properties of the planets. Coverage of a diverse stellar demographic will reveal previously unexplored planet populations. Moreover, with the timely second release of Gaia data (DR2) in April 2018 \citep{Gaia:2016, Gaia:2018}, the parent stars and the planets they host may be characterized better than ever before.

To exploit the scientific yield of the TESS mission, follow-up observations are expected to be a significant and essential part of the mission. Besides confirmation of the planetary nature of transit signals, follow-up studies will provide more complete characterizations of planetary systems, including, but not limited to, measurements of planet masses, orbital eccentricities, atmospheres \citep[e.g., chemical composition, albedo, temperature,][]{Seager:2010, Crossfield:2017}, and stellar obliquities \citep[e.g.,][]{Gaudi:2007, Albrecht:2012}. To date the majority of planets that have been characterized in detail with these methods are giant planets in close orbits around F, G, and K stars. TESS will provide the opportunity to answer questions such as: How diverse are the densities and compositions of close-in small planets? How diverse are the densities of planets formed in the same planetary system? How do planetary systems form around different types of stars?  What are the properties of planets with irradiation intensities similar to that of the Earth? The {\it Kepler} mission has provided hints of the answers to some of these questions \citep[e.g.,][]{Fulton:2017, Weiss:2014, Weiss:2018, Mulders:2015}. Given the expected improvement in the accuracy of planetary and stellar properties from TESS, we will be able to constrain some of these answers well enough to differentiate planet formation models.

A realistic estimation of TESS planet yields is expected to aid the efficient planning of follow-up observations, and to prepare the science community to use TESS results to address interesting scientific questions.

That is the subject of the work presented here. We carry out a simulation of TESS's planet yield using our best knowledge regarding the mission observation plan, current understanding of exoplanet populations, and the stars that TESS will observe. The simulation is applied to the two-year primary mission, as well as three scenarios of the TESS extended mission (yet to be funded by NASA). We highlight the following components of our planet yield simulation that are intended to make it more accurate than previous work \citep[i.e.,][]{Sullivan:2015, Bouma:2017, Barclay:2018}:

\begin{figure*}
    \centering
    \includegraphics[width=0.8\textwidth]{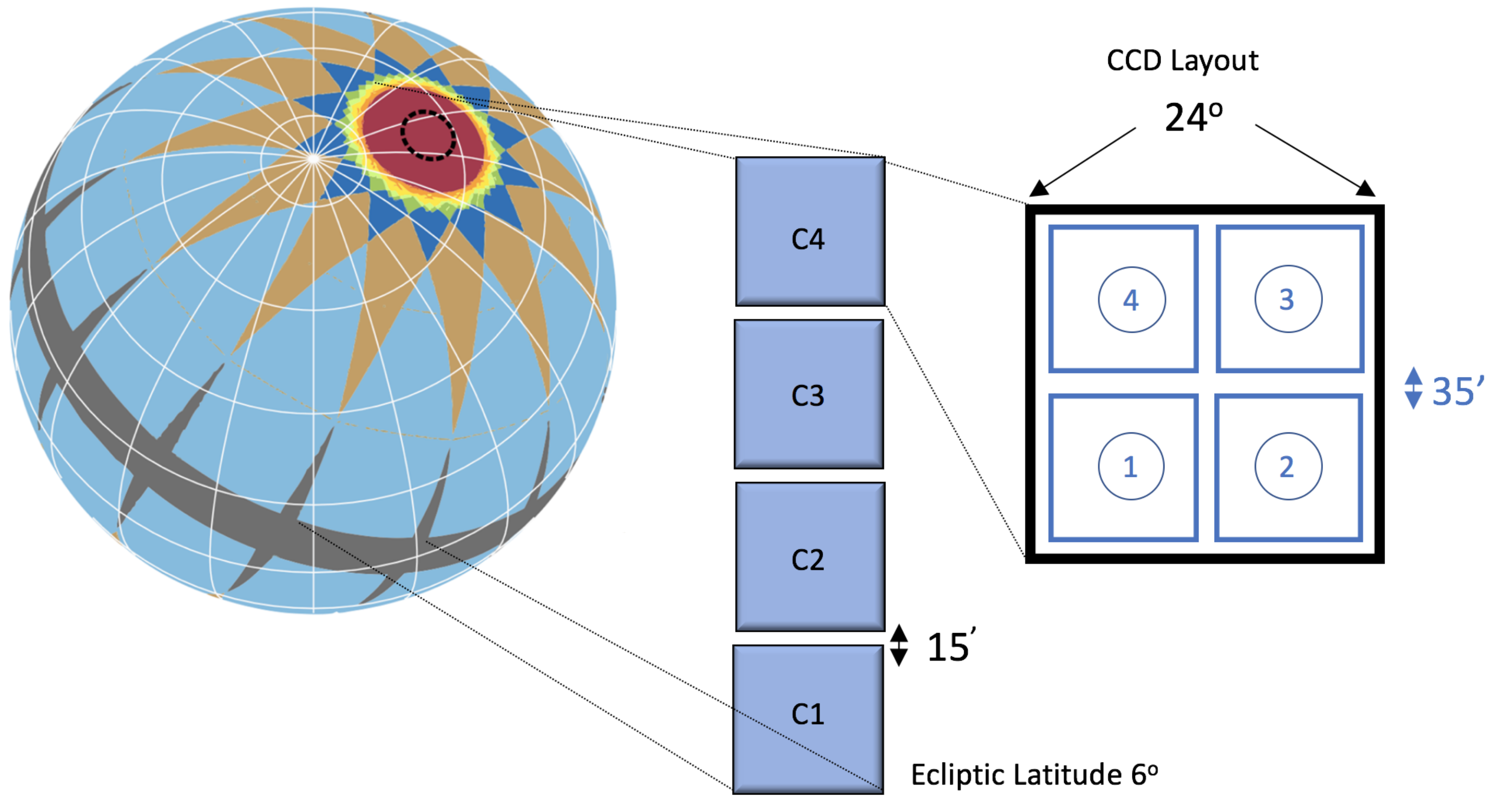}
    \caption{Schematics of the TESS field of view. The light blue area marks regions of the sky observed in a single TESS Sector, and areas marked in other colors are regions of the sky observed in more than a single sector. The area marked in red is observed continuously for 13 Sectors. That region is centered on the ecliptic pole and includes JWST continuous viewing zone marked in dashed black line. Camera \#1 (C1) is closest to the equator and Camera \#4 (C4) is centered on the ecliptic pole. The area in gray marks regions not observed by TESS, between ecliptic latitude of $-6$ deg and $+6$ deg, and in between neighboring sectors at low latitudes.
    The gaps between CCDs within the same camera are $\sim$ 100 pixels, leading to angular gaps of $\sim$ 35 arc minutes, and the angular gaps between fields of view of adjacent cameras is $\sim$ 15 arc minutes.
    Figure is partially adopted from \cite{Ricker:2014}.
    }
    \label{fig:sectors}
\end{figure*}

\noindent(1) Use of a photometric noise model that accounts for the nominal pointing jitter estimated through simulation prior to launch
\citep{Nguyen:2018}; 

\noindent(2) Reliable host star properties from the Gaia DR2 catalog \citep{Andrae:2018} for bright stars, and the TESS input catalog \citep[TIC,][]{Stassun:2017} for the other stars;

\noindent(3) A realistic process for selecting targets for the 2-minute cadence observations using the candidate target list \citep[CTL;][]{Stassun:2017} and information from Gaia DR2, while keeping the number of targets per CCD to within the maximum allowed by the mission;

\noindent(4) Simulation of multi-planet systems using a mutual inclination distribution based on multi-planet systems discovered by the {\it Kepler} mission \citep{Ballard:2016, Zhu:2018};

\noindent(5) Precise position and geometric distortion for the TESS field of view based on ray tracing results and pre-launch measurements.

We describe the simulation in detail in \secr{methods} and present the results for the TESS primary mission planet yield in \secr{results}. In \secr{extended} we describe three TESS extended mission scenarios, each lasting one or three years, and present the planet yield for each scenario. We conclude with a discussion and summary in \secr{discussion}.

\section{Methods}
\label{sec:methods}

\subsection{TESS field of view and observing strategy}

The TESS cameras, field of view, and observing strategy are described in detail in \cite{Ricker:2014, Ricker:2015} and \cite{Sullivan:2015}. We give here a brief description including the details most relevant to the present planet yield simulation.

TESS consists of four identical cameras each with an entrance pupil diameter of 10.5~cm and a field of view (FOV) of $\sim24\arcdeg \times 24\arcdeg$ (See Figure~\ref{fig:sectors}). The focal plane of each camera contains a mosaic of four 2058 pixel $\times$ 2048 pixel CCDs with a pixel scale of $\sim21\arcsec$ per pixel. 
The four FOVs are positioned adjacent to each other, forming a total FOV of $\sim96\arcdeg \times 24\arcdeg$. The angular gaps between camera fields are $\sim$15 arcminutes,  and within each camera the gaps between adjacent CCDs correspond to $\sim$35 arcminutes. These specific numbers describing a camera FOV are approximations used for convenience in the simulations. 

During the primary mission the set of four fields is oriented along a line of ecliptic longitude from $6\arcdeg$ below the ecliptic equator to $12\arcdeg$ past the Southern ecliptic pole in TESS's first year, and from $6\arcdeg$ above the ecliptic equator to $12\arcdeg$ past the Northern ecliptic pole in TESS's second year. Therefore the field viewed by Camera 4 will be centered on either the Southern or Northern ecliptic pole, and fully contain the corresponding JWST continuous viewing zone (See Figure~\ref{fig:sectors}).

\begin{figure}
    \centering
    \includegraphics[width=0.5\textwidth]{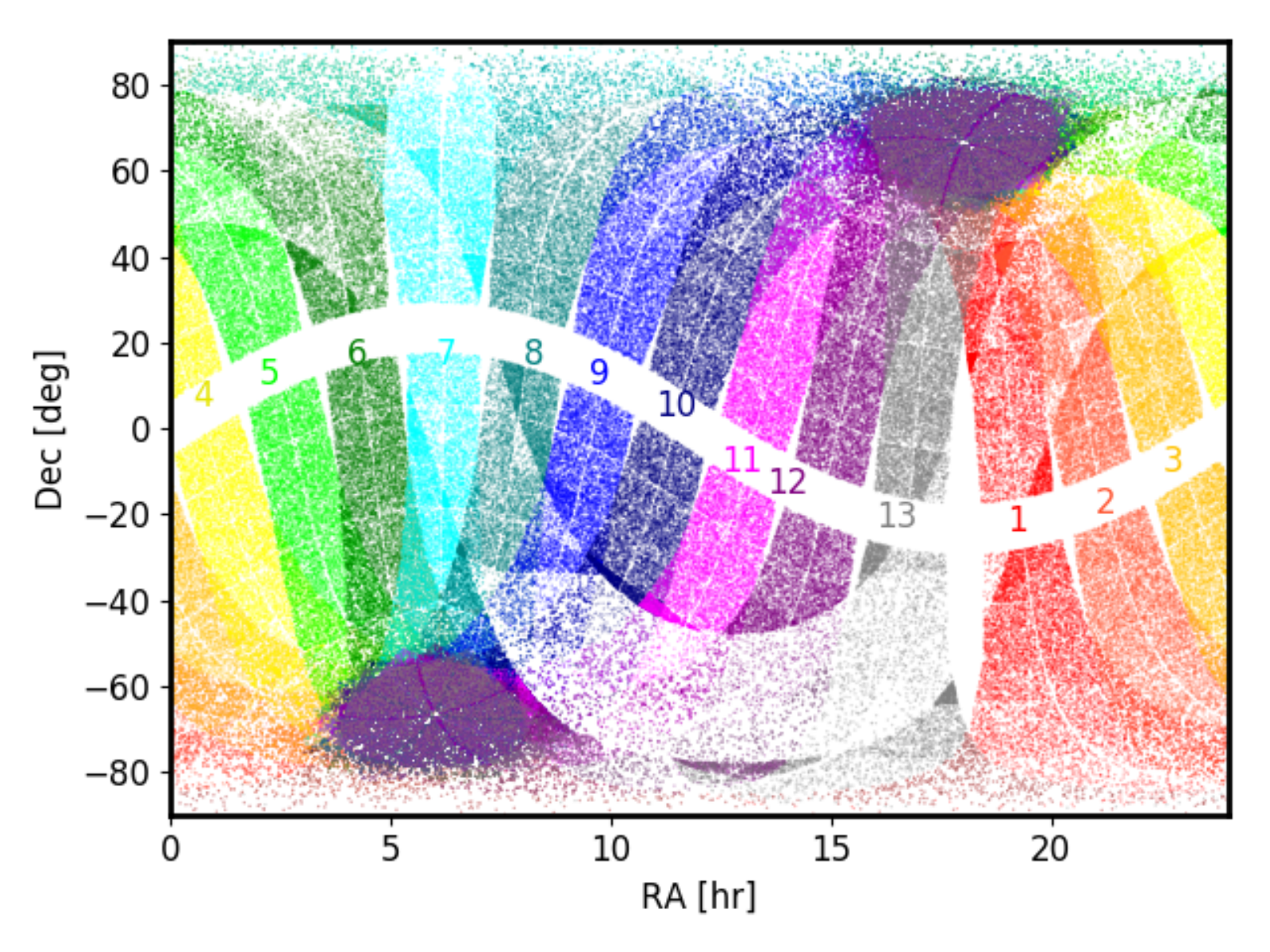}
    \caption{The footprint of different TESS sectors for the primary mission. The labeled numbers correspond to the sequence of the observations in both hemispheres. Each sector will be observed for 27.4 days, with the first sector starting from late June, 2018. The points represent a random draw of stars from different sectors (the density of points represent the density of 2 min cadence targets at a given position).}
    \label{fig:ra_dec}
\end{figure}

TESS is scheduled to start science observations of the first sector (Sector 1, or S1) in late June 2018 \footnote{TESS started its science observation on July 25, 2018.}. In each of the first and second years, TESS will collect observations from 13 different sectors with a cadence of 27.4 days. Every TESS camera will take an exposure every 2\,s. TESS will transmit to the ground mainly two types of data products made from the 2-s  images by stacking. The first data product will consist of 2-min stacks from limited regions around preselected target stars (target pixels, TP). The second product will consist of full images from all the cameras stacked at a 30-min cadence. The planned TESS fields of view for the primary mission are illustrated in equatorial coordinates in Figure \ref{fig:ra_dec}. We used the estimated mission pointing profile for the first year (Southern hemisphere) observations precomputed by the TESS team for the 18 April launch date and the nominal mission schedule \footnote{Actual sector positions will depend on the start date of science operations.}. The positions of the second year sectors are assumed to be symmetric about the ecliptic equator to the first year sector positions.

\begin{figure}[h]
    \centering
    \includegraphics[width=0.5\textwidth]{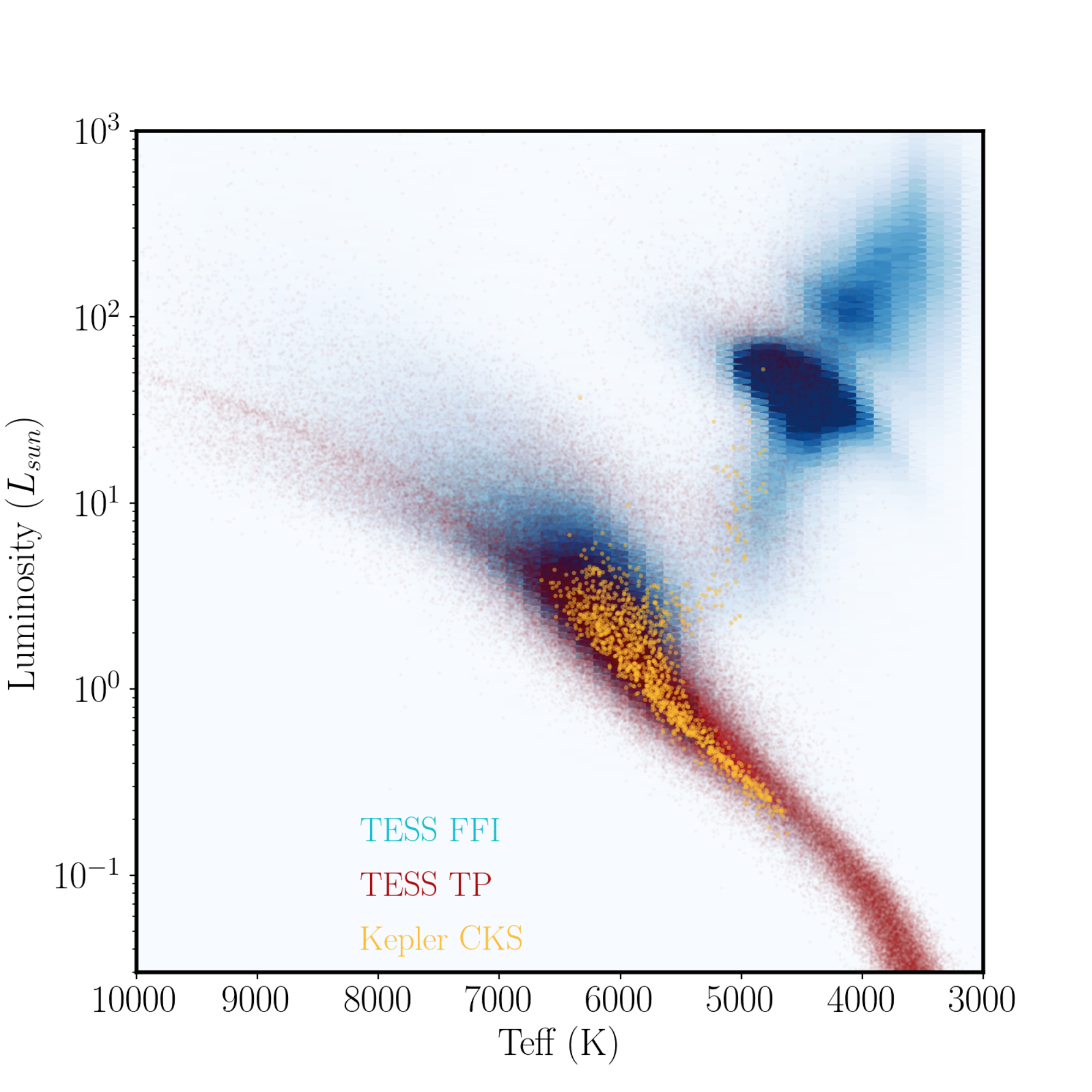}
    \caption{The HR diagram for stars TESS will observe. Blue dots represents stars with \tmag\ $<12$ mag observed in TESS FFIs (30-minute cadence), and red dots represent stars in the TESS target pixels (2-minute cadence). The stellar parameters are derived from Gaia DR2. For comparison, {\it Kepler} planet hosts with CKS parameters \citep{Petigura:2017} are represented with orange dots.}
    \label{fig:HR}
\end{figure}

\subsection{Star Catalogs}

\subsubsection{Target Pixel selection}

The targets for 2-min cadence target pixels (TPs) were selected from stars in the Candidate Target List (CTL) \citep{Stassun:2017} with TESS-band magnitudes \tmag$> 6$. We follow the mission requirement to limit the number of targets per CCD to 1,500, and the total targets per sector to 16,000. The target list is first populated with all the bright stars that fall on the science regions of the CCDs; then the list is reduced to satisfy the limitations on numbers by selecting targets from the CTL that have the highest priorities \footnote{the priority is computed based on the suitability of detecting transiting planets around the target (see \citet{Stassun:2017} for more details)}. When combing through the CTL we cross reference the star with Gaia DR2 parallaxes, if the star is indicated to be a giant in Gaia DR2, we do not include it in the 2 min cadence targets.    

We note these particular constraints result in approximately half the number of 2-min targets in Camera 4 (pointing at the ecliptic pole, i.e., JWST continuous viewing zone), compared to what was assumed in \citet{Sullivan:2015} and \citet{Bouma:2017}.

The preliminary TP list includes a total of 98,965 and 106,250 unique targets planned to be observed with 2-min cadence in the first and second year of operation, respectively.  

\subsubsection{Full Frame Images}

We selected $\sim$2.6$\times10^7$ stars in TIC-6 \citep{Stassun:2017} with catalog \tmag$< 15$ to be included in our Full Frame Image (FFI) simulation. The expected centroid locations of these target stars on the CCDs were predicted using a geometric model derived from prelaunch optical ray trace results \citep{Woods:2016}. Only stars that fall on science pixels are considered. For stars with \tmag$<$ 12, we use the information from Gaia DR2 \citep{Andrae:2018} to update the TIC-6 stellar parameters. The match between Gaia DR2 and TIC-6 catalog listings is based on a cross match using ecliptic coordinates\footnote{The matching accounts for proper motion based on Gaia DR2 (epoch 2015.5, equinox J2000).}, limiting the allowed difference between the Gaia ($G$) and TESS (\tmag) magnitudes, as well as limiting the allowed difference between corresponding estimated values of $T_{\rm eff}$. A more thorough incorporation of the Gaia DR2 catalog into the CTL and TIC is an ongoing task of the TESS Target Selection Working Group, and is beyond the scope of this paper. The typical uncertainty on the stellar radius from Gaia DR2 for our sample is $\sim$10\%, which is comparable to the typical uncertainties from high-resolution optical spectra (combined with evolutionary models) obtained using Keck/HIRES for the {\it Kepler} planet host star sample \citep{Petigura:2017}. 

The Gaia DR2 stellar parameter estimation module imposes a lower limit on the stellar radius of 0.5 $R_{\rm \odot}$, which means the stellar radii from Gaia DR2 are not suitable for a large fraction of the M dwarf population. Therefore, for stars with TIC $T_{\rm eff}$ below 4,000~K we used the Gaia DR2 parallax to determine if the star is an M dwarf or an evolved giant star by estimating the stellar radius using the target's distance and observed brightness. We then re-estimate the radius and mass of the M dwarfs following empirically calibrated model-independent relations between \teff and radius \citep{Mann:2015}.

\figr{HR} shows the distribution of stars observed by TESS in the HR diagram. Compared with {\em Kepler}, TESS will survey stars with a much broader spectral type distribution, hopefully detecting the planets of host stars spread throughout most of the HR diagram. 

\begin{figure*}
    \centering
    \includegraphics[width=\textwidth]{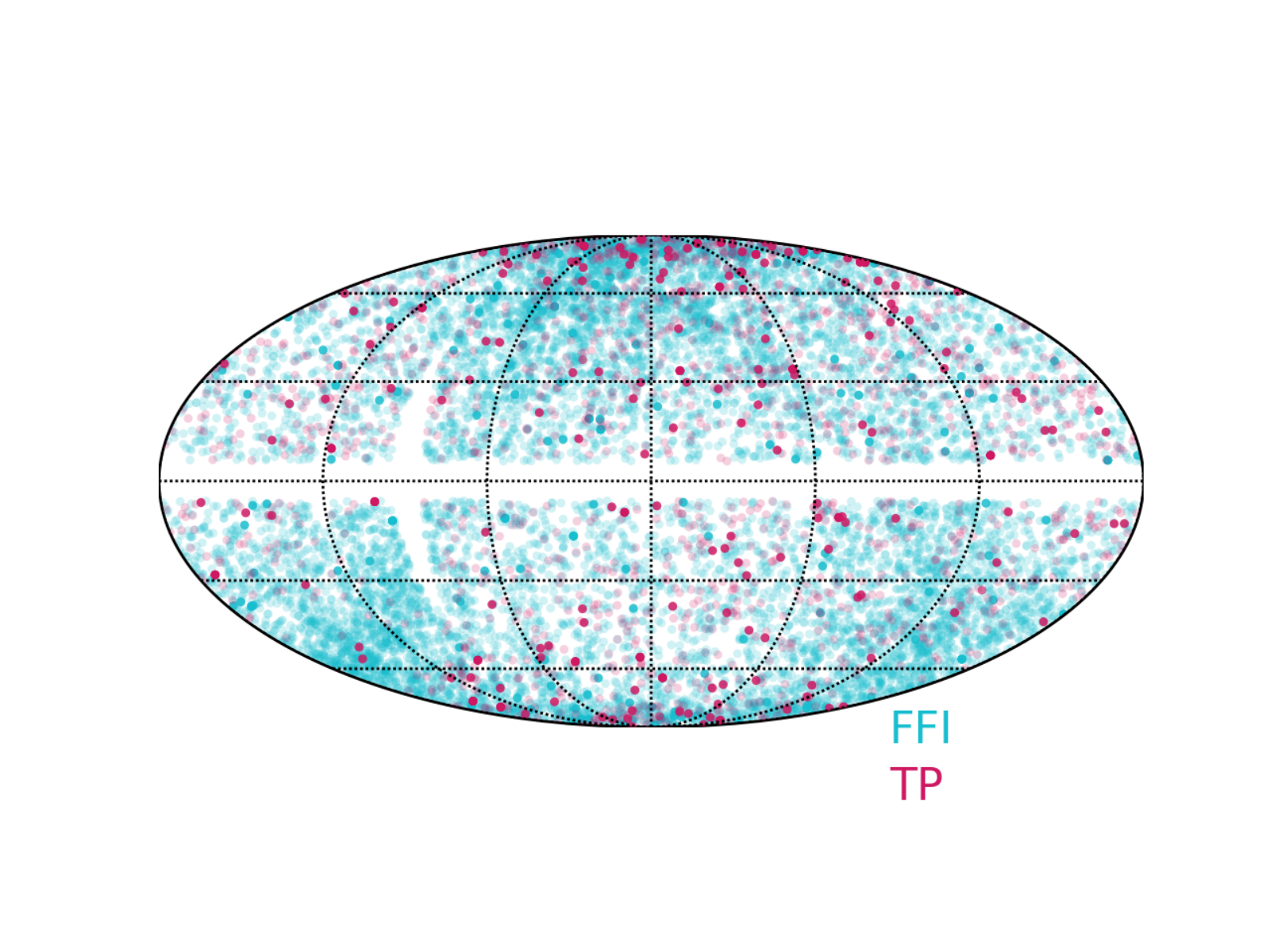}
    \caption{Expected yields of planets in ecliptic coordinates for the primary TESS mission. Cyan dots are planets detected from FFIs, red dots are planets detected from target pixels. Darker color represents star systems with higher numbers of transiting planets detected by TESS. We note that the sky within 6$^\circ$ of the ecliptic plane is not observed.}
    \label{fig:gc}
\end{figure*}

\subsection{Planet Yield simulations}

Our estimation of the planet population follows that of \citet{Sullivan:2015}. We adopted a planet occurrence rate for M dwarfs (\teff$\,<\,4000\,K$ and $\log\,g > 3$) per \citet{Dressing:2015}; for other stars, we adopted a planet occurrence rate per \citet{Fressin:2013}. For large planets, in the largest radius bin of \citet{Fressin:2013}, 6--22 $R_E$, we re-sampled the radius within the bin boundaries from a Gaussian distribution centered on 1 $R_J$, with a width of 0.2 $R_J$.

The signal to noise ratio of the transit signal of a planet in a single TESS sector $s$ is calculated using: 
\begin{equation*}
    {\mathrm SNR}_s = N_{\rm tran}^{1/2}  \delta\times((\sigma_{hr}/T_{23,h})^2 + \sigma_v^2)^{-1/2}\,(1+f)^{-1},
    \label{eq:snr}
\end{equation*}
in which, $N_{\rm tran}$ is the number of transits per sector, the transit depth $\delta = (R_p/R_*)^2$,  $T_{23, h}$ is the transit duration in hours between the second and third contact points, $f$ is the contribution of flux from neighboring stars relative to the total flux of the target star measured in the photometric aperture. For M dwarfs, we estimate the noise contribution from stellar variability $\sigma_v$ over a 1-hour timescale by drawing from a log uniform distribution in the range 20-500 ppm. 
This is motivated by Figure 7 in \citet{Sullivan:2015}. For other stars, we assume $\sigma_v = 0$, since their astrophysical variability amplitudes are typically smaller than the noise floor of TESS. 
We use the updated photometric precision estimates of \citet{Nguyen:2018} for the expected TESS noise level over a 1-hour timescale $\sigma_{\mathrm hr}$. Similar to \citet{Sullivan:2015}, \citet{Nguyen:2018} created synthetic images using model TESS point spread functions (PSFs) for different field angles \footnote{In total, photometric noise is modeled as a function of \tmag\ at 5 different field angles in the simulation.}. In addition, \citet{Nguyen:2018} incorporated in the simulation the effect of spacecraft pointing jitter estimated on the basis of a  prelaunch TESS fine pointing simulation. The overall noise performance is comparable to \citet{Sullivan:2015} Figure 14, except that the systematic noise floor from jitter is around 40 ppm (rather than 60 ppm) for the bright stars (\tmag $\sim$ 7), and slightly higher for the fainter stars. For each star, we first estimate the location of the star on TESS CCDs for each observation sector separately, and then adopt the noise model corresponding to the relevant portion of the detector.
For targets observed in more than a single sector, we quadratically combine the signal-to-noise of the target from each observed sector to calculate the final signal-to-noise ratio (SNR) at a given time.  
We require a detected planet to have a SNR greater than 7.3 and at least two transits recorded by TESS.      

\begin{figure*}
    \centering
    \includegraphics[width=\textwidth]{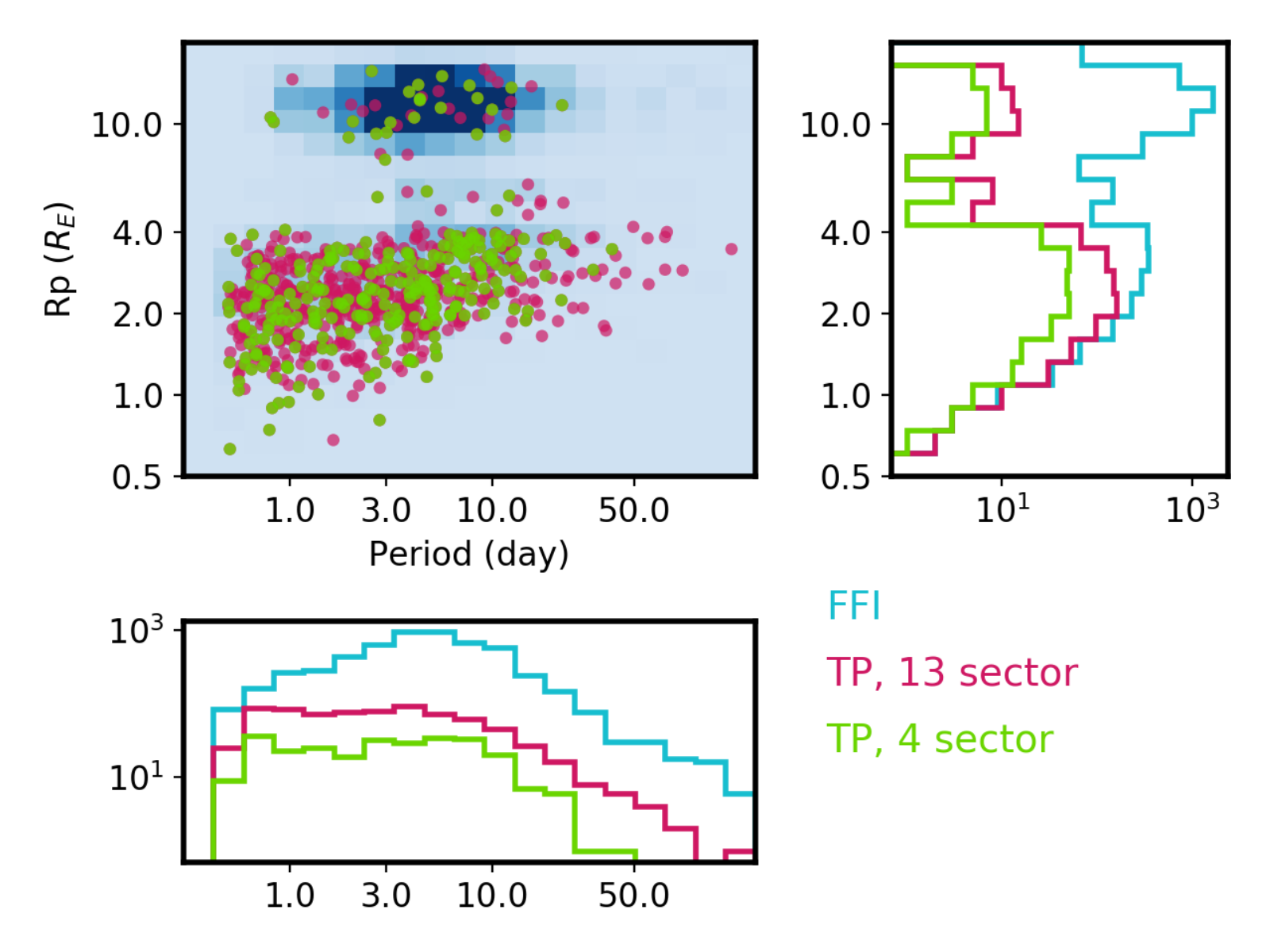}
    \caption{The expected period and radius distribution of planets around stars with \tmag$<15$ mag discovered in TESS first year of operations. The cyan background shows the 2-d histogram of planets discovered in the FFIs.
    The green, and red dots represent planets discovered in the target pixels after 4, and all 13 sectors of observations.}
    \label{fig:Rp_P}
\end{figure*}

\subsubsection{Multiple planets}

Our experience with the {\it Kepler} mission and ground based radial velocity surveys \citep{Wright:2009} has taught us that multi-planet transiting systems are abundant \citep{Latham:2011, Lissauer:2011, Fabrycky:2014, Tremaine:2012}. When accounting for multiplicity, previous studies typically simulate planets in the same system with independent occurrence probabilities, and assume the system to be perfectly co-planar \citep{Sullivan:2015, Bouma:2017, Barclay:2018}. However, studies of {\it Kepler} planetary systems have revealed that there likely are correlations between the mutual inclination distributions and the true planetary multiplicities \citep{Johansen:2012,HM13, Ballard:2016, Zhu:2018}. To more realistically investigate the multi-transiting planet systems that may be discovered by TESS, multiple planetary systems are injected into our simulations based on updated information on the dependence of a system's mutual inclination distribution on the number of planets hosted by the target star. For stars other than M dwarfs, we follow the approach of \citet{Zhu:2018} when the first planet in the system is not a hot Jupiter (a hot Jupiter is defined here as $R_p>6\,R_E$ and $P<10$\, day)\footnote{Since the companion occurrence rate in hot Jupiter systems is not well constrained.}. The probability of a target star hosting a planetary system is 30\%, while the probability of a star hosting $N$ planets ($N=$\,1--6) is evenly distributed\footnote{This is an approximation to the fitting result of the probabilities for each intrinsic multiplicity category fitted in \citet{Zhu:2018}}. The systems hosting higher multiplicity are more likely to be co-planar, with the mutual inclination dispersion following a Rayleigh distribution in which the Rayleigh width is $\sigma_i = 20^{\circ}\,N^{-2}$. For intrinsic single planet systems, the orientation of the planet's orbit is selected at random.  

For M dwarfs, we follow the simpler ``Kepler dichotomy" approach of \citet{Ballard:2018}. The probability of a target star hosting a single planetary system is 85\%. For multiple planetary systems, we used the best fit value for {\it Kepler} M-dwarf systems from \citet{Ballard:2016}, with the average number of planets being $N=4.6$, and the Rayleigh width of the mutual inclination distribution being $\sigma_i = 1.4^{+6.3}_{-1.2}\,^{\circ}$. With a relatively small mutual inclination dispersion, the transit probability of a second planet is enhanced compared to the co-planar assumption, given the inner planet also transits. \citet{Ballard:2018} reported an enhanced planet yield by adding the mutual inclination distribution to the M-dwarf samples of \citet{Sullivan:2015}. We require that the multiple planetary systems follow the \citet{Deck:2013} stability criterion:
$\frac{P_{\mathrm out}}{P_{\mathrm in}} > 1.16$.

\begin{table}[]
    \centering
    \caption{Primary Mission Estimated Planet Yields\label{tab:yield}}
    \begin{tabular}{l|r|r|r}
    \hline\hline
        Magnitude limit & 10 & 12 & 15 \\
     \hline
       FFI AFGK & 1176 & 2983 & 8591\\
        \hline
        FFI M\footnote{The FFI planet injection simulation is done independently from the TP injection simulation.} & 96 & 695 & 2260\\
        \hline
        TP AFGK & 276 & 451 & 453\\
        \hline
        TP M &125 & 569 & 1337\\
        \hline\hline
    \end{tabular}
\end{table}


\section{Results}
\label{sec:results}

\subsection{Primary Mission}

Table \ref{tab:yield} presents the yield of planets from the TESS primary mission. For the $\sim2\times10^5$ unique stars on target pixels, we expect to detect $\sim$ 2000 planets. The FFIs are expected to provide another $\sim 10^4$ detected planets. We show the sky positions of the detected planets in Figure \ref{fig:gc}. 
Stars with a higher number of detected transiting planets are mostly located close to the ecliptic pole, where the observational baseline for an average star is the longest. 
    Figure \ref{fig:Rp_P} shows the distribution of periods and radii of the planets detected in the target pixels and the FFIs. Specifically, we examined the expected yields of planets from the target pixels after 4 and after 13 sectors of observation. More than 50 Level 1 \footnote{The Level One Science Requirement for the TESS Mission is to measure masses for 50 transiting planets smaller than 4 Earth radii} planets ($R_p < 4 R_E$) are expected to be detected in the first four science sectors\footnote{The first TESS public data release will be the first four science sectors}. As the data from each new sector accumulates, TESS will deliver smaller planets with longer orbital periods. 

\begin{figure}
    \centering
    \includegraphics[width=0.5\textwidth]{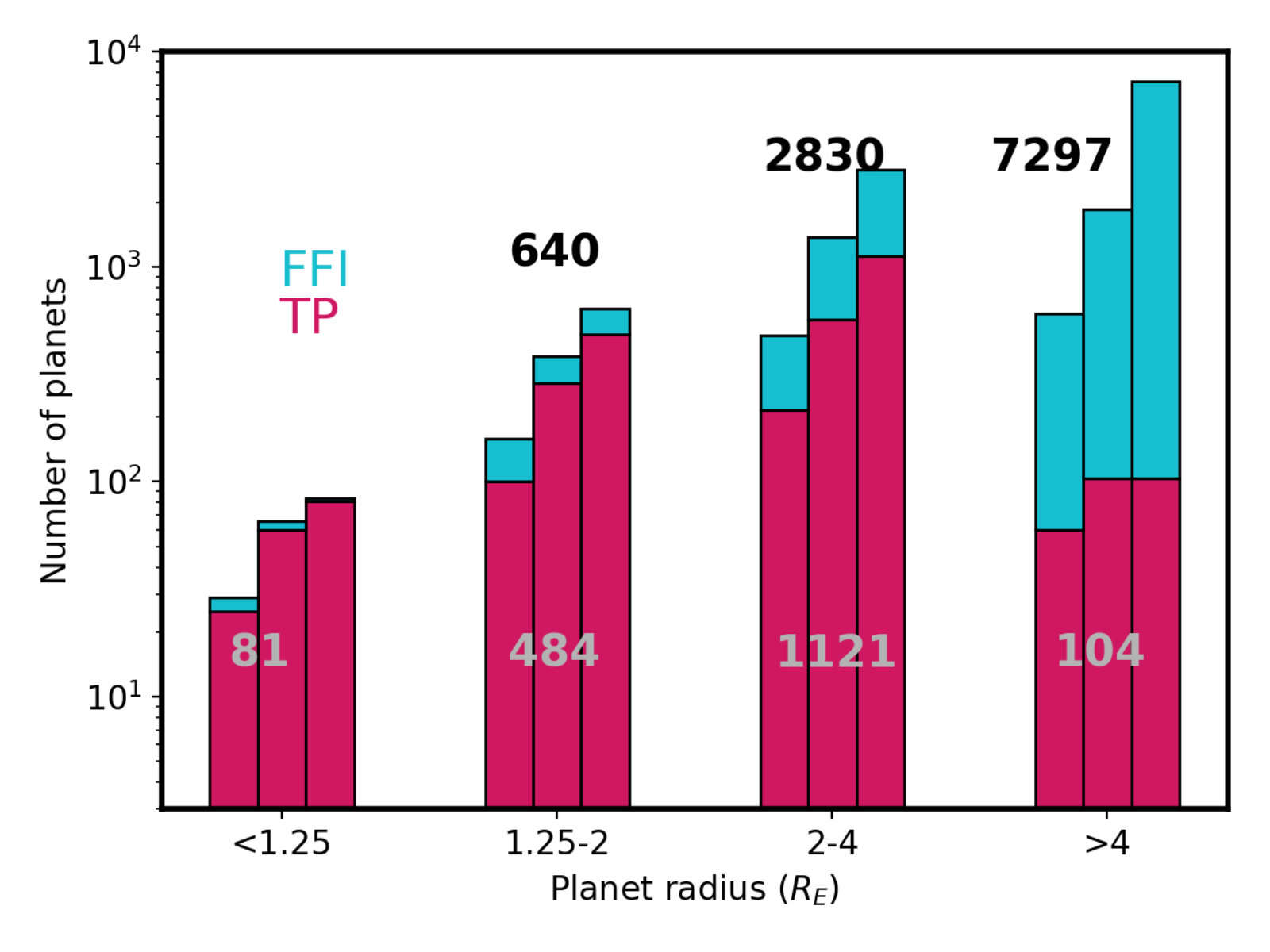}
    \caption{The expected yield of planets in the TESS 2-year primary mission. The three different bins represent stars brighter than \tmag\ of 10, 12, and 15 mag. Red represents planets discovered in target pixels, and cyan represents planets discovered in FFIs only. The numbers labeled on top of each planet category shows the number of discoveries for stars with \tmag$\,<15$ mag. Grey numbers represent the planets discovered in target pixels only, and the black numbers represent the total.}
    \label{fig:HistTotal}
\end{figure}

Figure \ref{fig:HistTotal} shows the number of planets TESS is expected to discover in the primary mission in different size categories. The majority of the smallest planets ($R_p<2R_E$) from TESS discoveries will have both target pixel and FFI observations. This will ensure a more accurate determination of the transit ephemerides, and benefit follow up observations. 
Among the TESS discoveries, $\sim$1000 planets will be around stars with \tmag $<$ 10, which allows detailed characterization of planet properties using ground based facilities. 
The uncertainties on the planet yield estimation comes mainly from three sources: (1) Poisson statistical variations; (2) uncertainties in occurrence rates and our multiplicity model;  and (3) the uncertainties in the stellar radii, especially for the late M dwarfs. Our planet yield estimate is comparable to previous works \citep{Sullivan:2015, Bouma:2017, Barclay:2018}. 
\citet{Sullivan:2015} used stars from galactic models, their yields are similar to our \tmag $<15$ sample, especially for planets smaller than $4\,R_E$. We report a smaller number of discovered giant planets. This is because the \citet{Sullivan:2015} FFI sample included stars slightly fainter than \tmag$=15$ and TESS is able to discover giant planets beyond the magnitude limit in our simulation. \citet{Bouma:2017} and \citet{Barclay:2018} only simulated relatively bright stars: the yields from their simulations are between our \tmag$<12$ mag and \tmag$<15$ mag sample yields, as expected. We also compared our planet yields for M dwarfs with \citet{Muirhead:2018}. Their 50,000 brightest stars scenario is similar to a scenario with a magnitude limit of \tmag$=13$ for M dwarfs. We predict $\sim1296$ planets from this sample, which is about 1.4 times higher. This enhancement in total number of planets is mostly due to our different assumptions relative to the occurrence rates of multiple planetary systems (see \citealt{Ballard:2018}).

\begin{figure}
    \centering
    \includegraphics[width=0.5\textwidth]{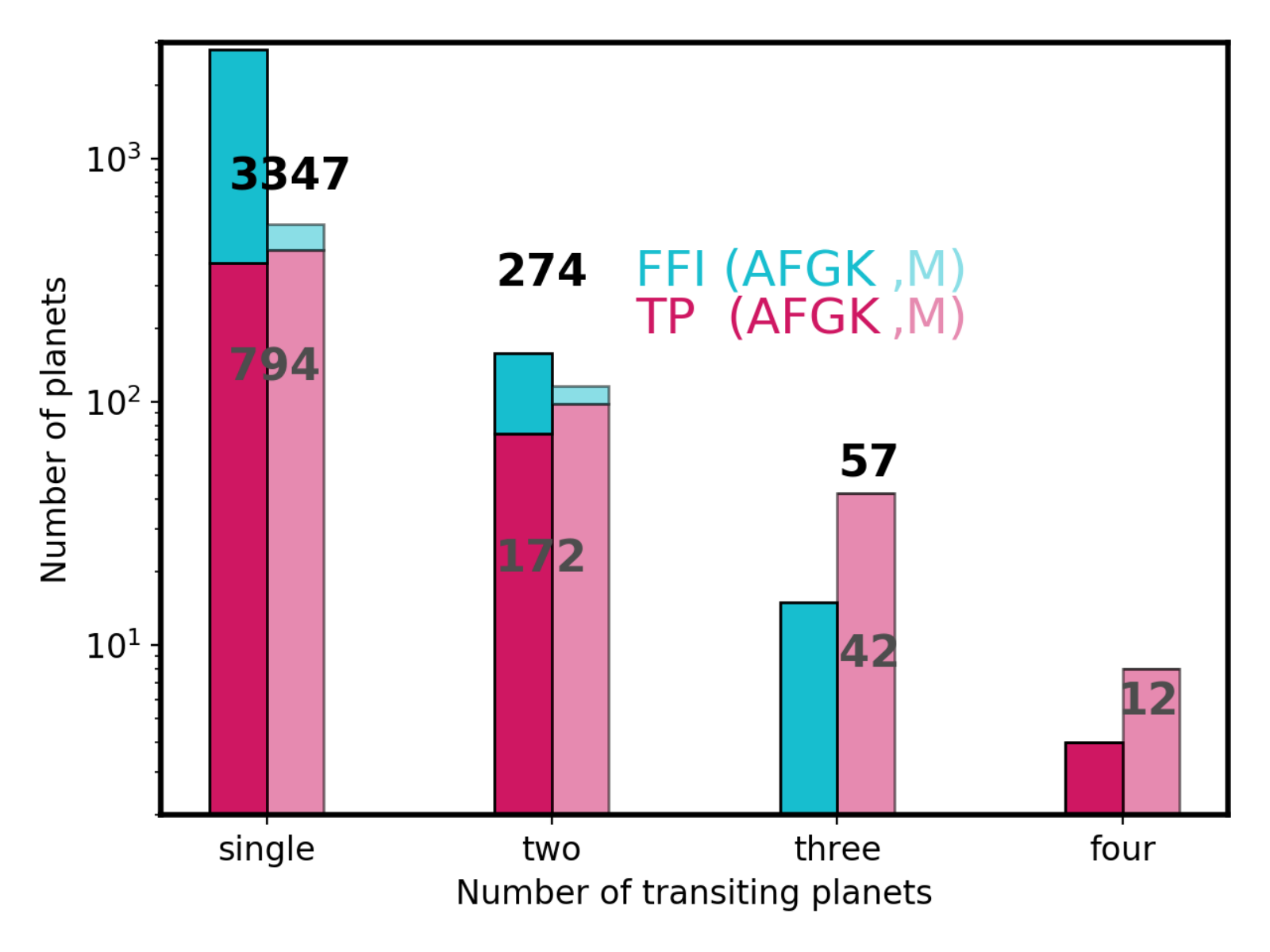}
    \caption{The expected yields of planets in multi-transiting planet systems around stars with \tmag\ brighter than 12 mag in TESS 2-year primary mission. Red represents planets discovered in target pixels, and cyan represents planets discovered in FFIs only. The darker bins represent planets around A, F, G, and K stars; the lighter bins represent planets around M dwarfs. The labeled numbers show the sum of all planets in each category.}
    \label{fig:HistMulti}
\end{figure}
 
\begin{figure}
    \centering
    \includegraphics[width=0.5\textwidth]{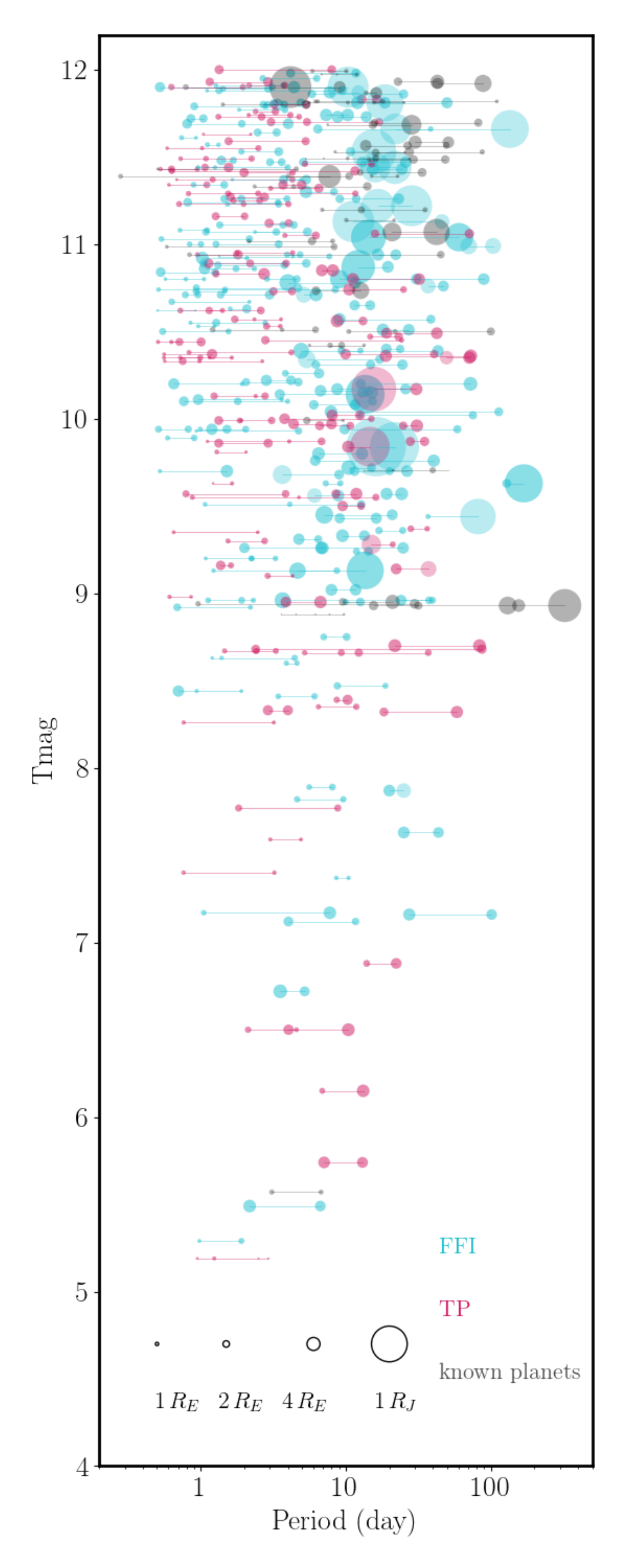}
    \caption{A demographic view of the expected yields of multi-transiting planet systems around stars with \tmag\ brighter than 12 from the primary mission of TESS. Red represents planets from target pixels, cyan represents planets from FFIs only. For comparison, we show in gray the currently known multi-transiting planet systems. The radius of the dots are proportional to the radius of the planets (see the top of the chart for scale references). Planets orbiting the same host star are connected with a horizontal line. 
    \label{fig:Multi}}
\end{figure}

TESS will also discover a few hundred systems in which a single star hosts multiple transiting planets. 
For the primary mission, we expect $10\%$ of the planetary candidates to be in multiple transiting planetary systems. For M dwarfs specifically, $\sim\,15\%$ of the discovered system will have multiple transiting planets, consistent with the estimate of \citet{Ballard:2018}.

We show the architecture of all predicted multiple transiting planet systems with \tmag $<$ 12, with known planets for comparison, in Figure \ref{fig:Multi}. Though relatively few are known at present, TESS will identify hundreds of new multiple transiting systems around bright stars. Some of these systems will enable follow up observations to measure individual planet densities, atmospheres, obliquities, and eccentricities within the same system, leading to a better understanding of planet formation and evolution. 

We expect that TESS will observe, during the primary mission, just one transit event for each of hundreds of longer-period transiting planets \citep{Villanueva:2018}. We conservatively use a higher detection threshold for these events (SNR $>$ 10), and find that 75 and 689 of the single-transit events will be caused by true planets in the target pixels and FFIs, respectively. If they can be confirmed, these single-transit planets will uniquely expand certain exoplanet research avenues and even open new ones. These planets will extend the period range over which it will be possible to study the atmospheres and mass-radius relation of exoplanets, especially for FGK stars. If all the single-transit planets that are temperately irradiated ($0.32 < S/S_E < 1.78$; \citealt{Kopparapu:2013}) and have FGK dwarf star hosts can be confirmed, the number of such temperate planets will grow from 5 (that are expected to transit more than once in TESS light curves) to 16. Finally, while we did not explicitly include circumbinary planets (CBPs) in our yield simulations (so we do not have {\it absolute} numbers for their yield), we expect that the number of single-transit CBPs will be larger than the number of CBPs that TESS would otherwise find by at least a factor of a few \citep{Quarles:2018}.

\section{Extended Mission}
\label{sec:extended}

\begin{figure*}
    \centering
    \includegraphics[width=\textwidth]{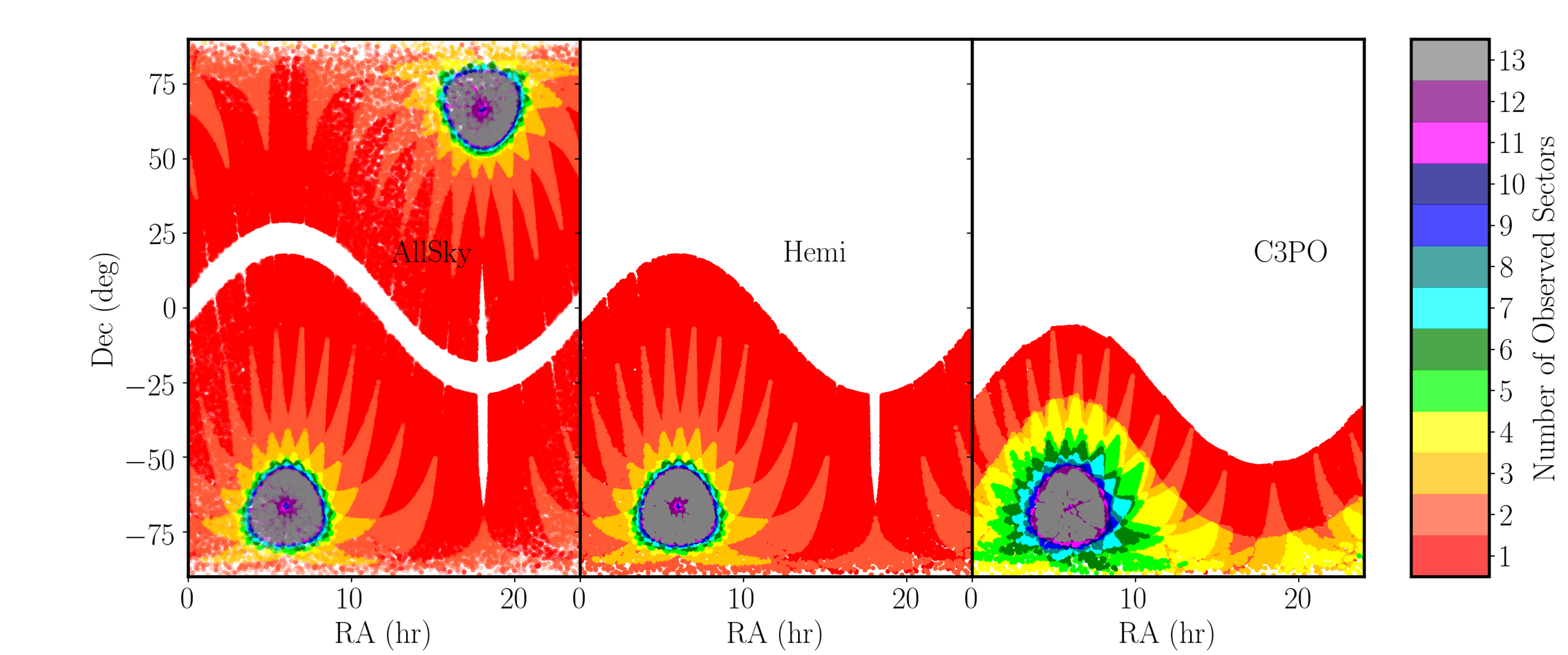}
    \caption{TESS sky coverage of Extended Mission for the assumed third year operation pointing model. The {\bf Allsky} model will have half the baseline for each sector compare to the other models. We note that the {\bf Hemi} and {\bf C3PO} scenario are only investigated for the southern hemisphere. An equivalent survey strategy for the northern hemisphere should produce similar results.
    \label{fig:EM_Sky}}
\end{figure*}

\begin{figure*}[h]
    \centering
    \includegraphics[width=\textwidth]{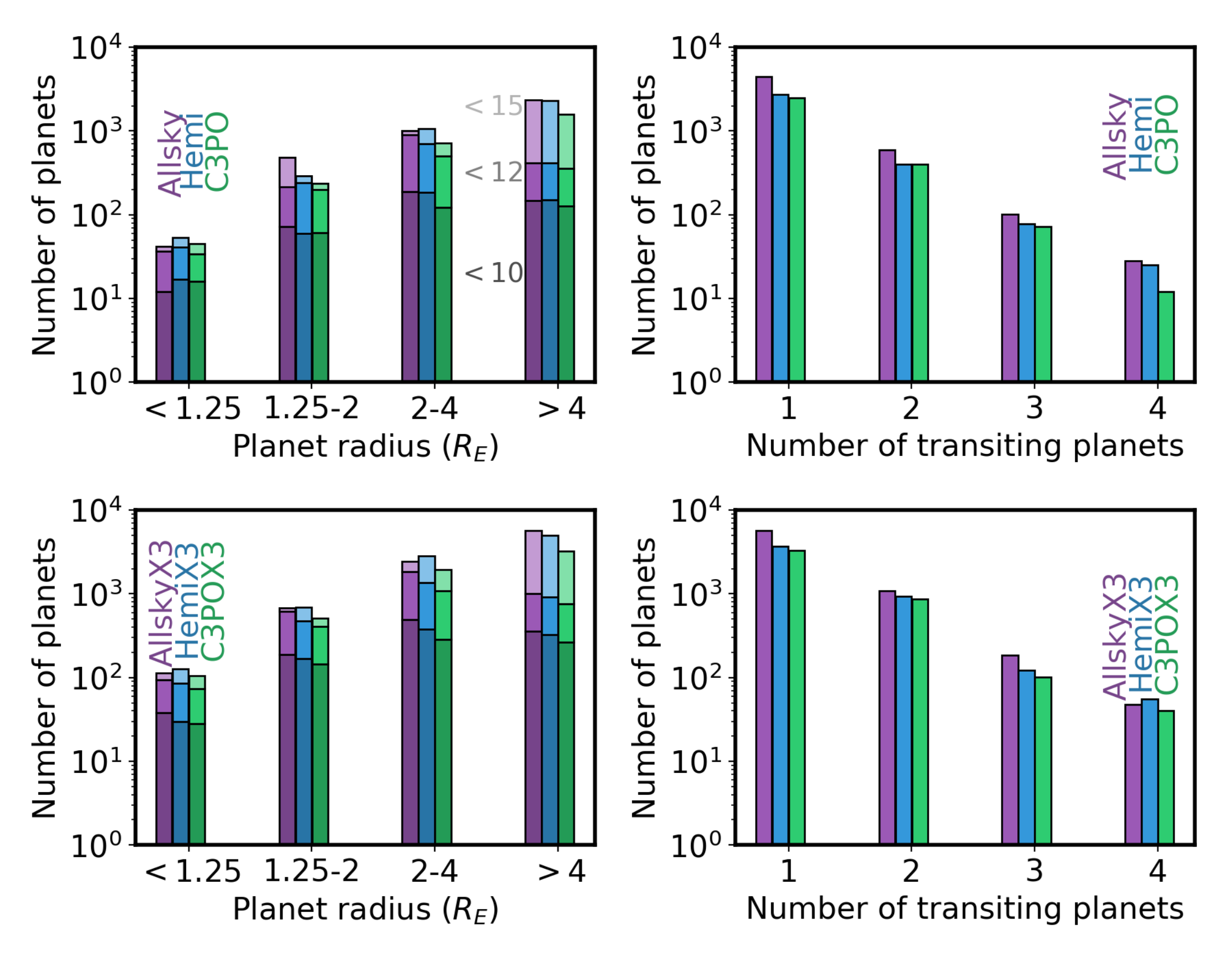}
    \caption{Top Panel-Left: expected planet yields from a third year of operation with different extended mission pointing models. For each pointing model, the short, medium and tall bars represent planets around stars brighter than 10, 12 and 15th magnitude. Top Panel-Right: expected number of planets in multiple transiting planet system around stars brighter than 12th mag at the end of year 3 of the entire mission. Bottom Panel: similar to top panel, but for the expected yield with 3 years extended mission repeating the first year extended mission scenario.
    \label{fig:EM_yields}}
\end{figure*}
  
\subsection{Extended mission pointing scenarios}

We investigated the following three pointing scenarios for the third year of TESS operations. The first two scenarios were also described in \citet{Bouma:2017}, for which we re-evaluate the outcomes with our updated simulations. We did not evaluate the Pole scenario from \citet{Bouma:2017} because that scenario brings one of the camera fields (Camera 1 or Camera 4) too close to the Sun. Instead we study a scenario that centers Camera 3 on the ecliptic pole, which follows a similar motivation of extending the temporal baseline in the ecliptic pole region.  
We detail the three scenarios below:

{\bf Hemi}: The third year of operation is an exact repeat of the southern hemisphere pointings from the first year. 

{\bf Allsky}: The spacecraft alternates observing the north and south hemisphere, with each field observed for 13.7 days instead of 27.4 days as in the primary mission. 

{\bf \asypole}: The spacecraft centers its camera 3 on the southern ecliptic pole, keeps its pointing in galactic latitude fixed, and changes its pointing in galactic longitude evenly between sectors.

The sky coverage of each pointing scenario is illustrated in Figure \ref{fig:EM_Sky}. The {\bf Allsky} scenario has the most sky coverage but the shortest average baseline observation duration per star, while the {\bf \asypole} scenario is intended to achieve the opposite.   

We performed a similar planet yield simulation for each of the three extended mission pointing scenarios. For planets with orbital periods exceeding the period range in the occurrence rate table we used for the primary mission simulation, we extrapolate their occurrence rate to periods of $\sim$2,000 days assuming the planet occurrence rate follows a log uniform distribution at long periods. The results for the FFIs are shown in Figure \ref{fig:EM_yields}. The total number of new planets expected to be discovered in the third year is of order of 2,000, with a significant number of those hosted by bright stars. The total number of new planets, as well as the multiplicity distribution at the end of the third year do not differ significantly among the three scenarios (see Figure \ref{fig:EM_yields}). However, the \asypole\ pointing yields more planets with P$>$50 days in all radius classes compared to other scenarios. This statement holds when we look at stars with \tmag\ brighter than 12. We do not attempt to simulate planet yields from the TPs for the extended mission, since the selection criterion of TPs may evolve significantly from the strategy we assumed for the primary mission simulation. 

\begin{figure}
    \centering
    \includegraphics[width=0.5\textwidth]{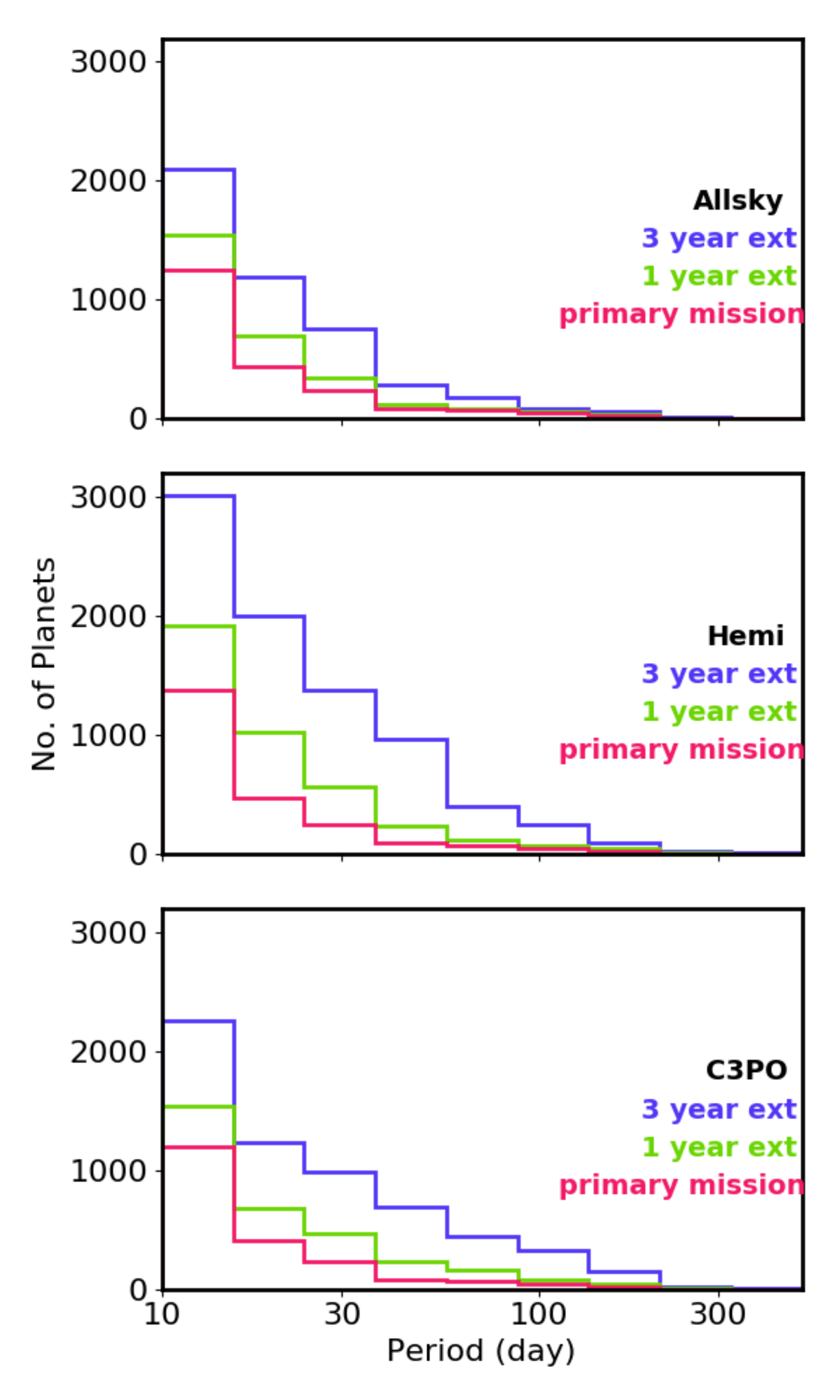}
    \caption{Expected yields of planets with $P>10\,$day for a TESS extended mission (1yr and 3 yr) with three different pointing scenarios. From the top to bottom panel: we show the result for {\bf Allsky}, {\bf Hemi} and \asypole\ scenario. We assume the 3 year extended mission uses the same pointing scenario every year. Red, orange, blue histograms (stacked on top of each other) represent planets discovered in the TESS primary mission, a one year extended mission, and three year extended mission.  
    \label{fig:Ext_All}}
\end{figure}

Given an extended mission that will last three years, we evaluate the planet yields for repeating the pointing from the third year of operation for three years. 
The {\bf Hemi} and \asypole\ scenarios have a slight advantage over {\bf Allsky} since the relatively long baseline allows these two scenarios to discover planets with longer periods and smaller transit depths (Figure \ref{fig:Ext_All}). To be specific, the \asypole\ , {\bf Hemi} and {\bf Allsky} scenarios discover $\sim$ 300, 200 and 100 planets with \tmag $<12$ and $P>50$\,days. Depending on the scientific priorities of a TESS extended mission, one could anticipate a permutation of the above three scenarios over the three years. Additional investigation is a subject of ongoing work by the TESS Science Team, and beyond the scope of this work. Figure \ref{fig:Ext_Flux} shows that a TESS extended mission will discover a number of small planets ($R<2R_E$) that are temperately irradiated and that orbit larger stars compared to the discoveries likely in the primary mission (we use the \asypole\ scenario as an example in that figure). 

\begin{figure}
    \centering
    \includegraphics[scale=0.283]{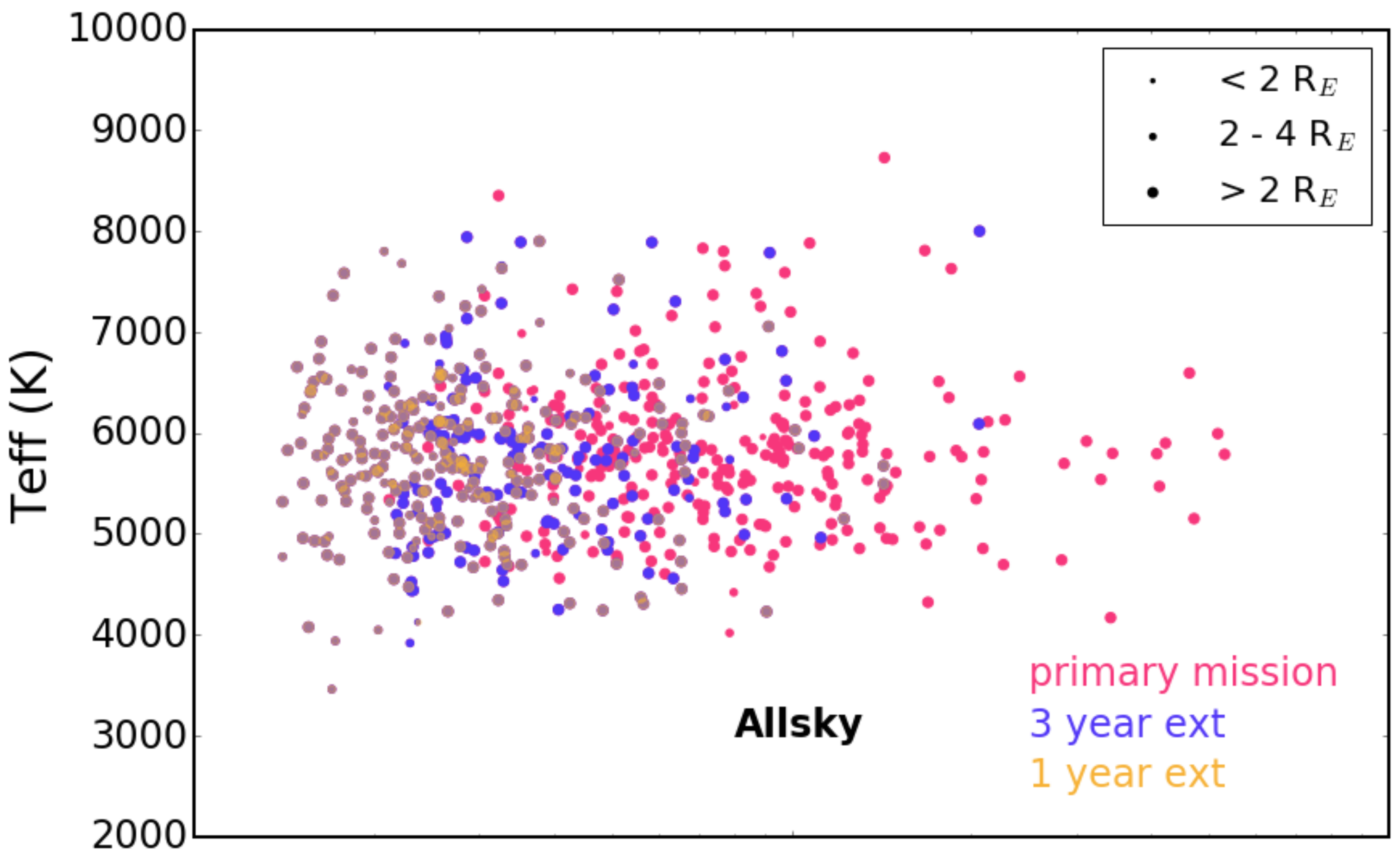}
    \includegraphics[scale=0.283]{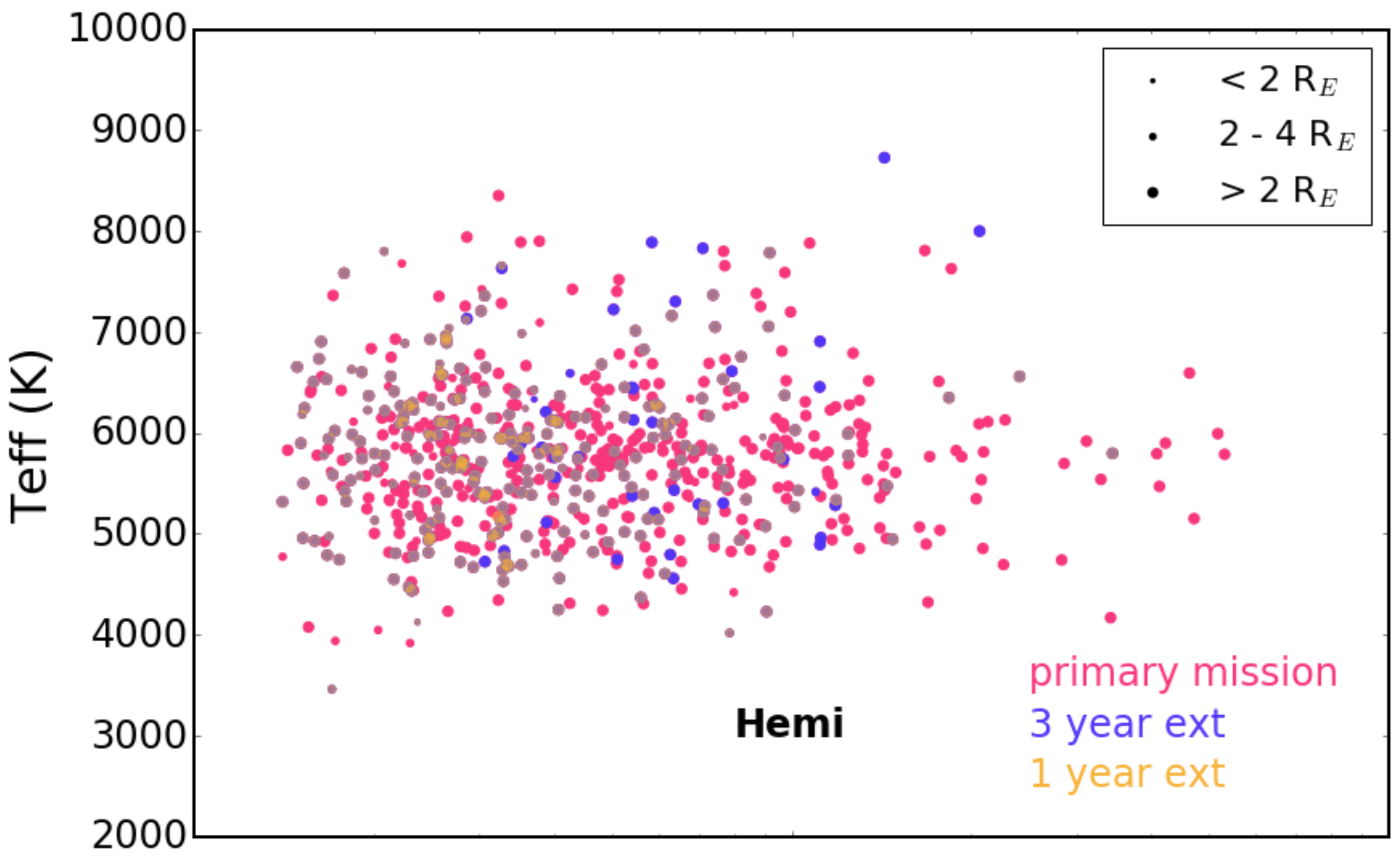}
    \includegraphics[scale=0.283]{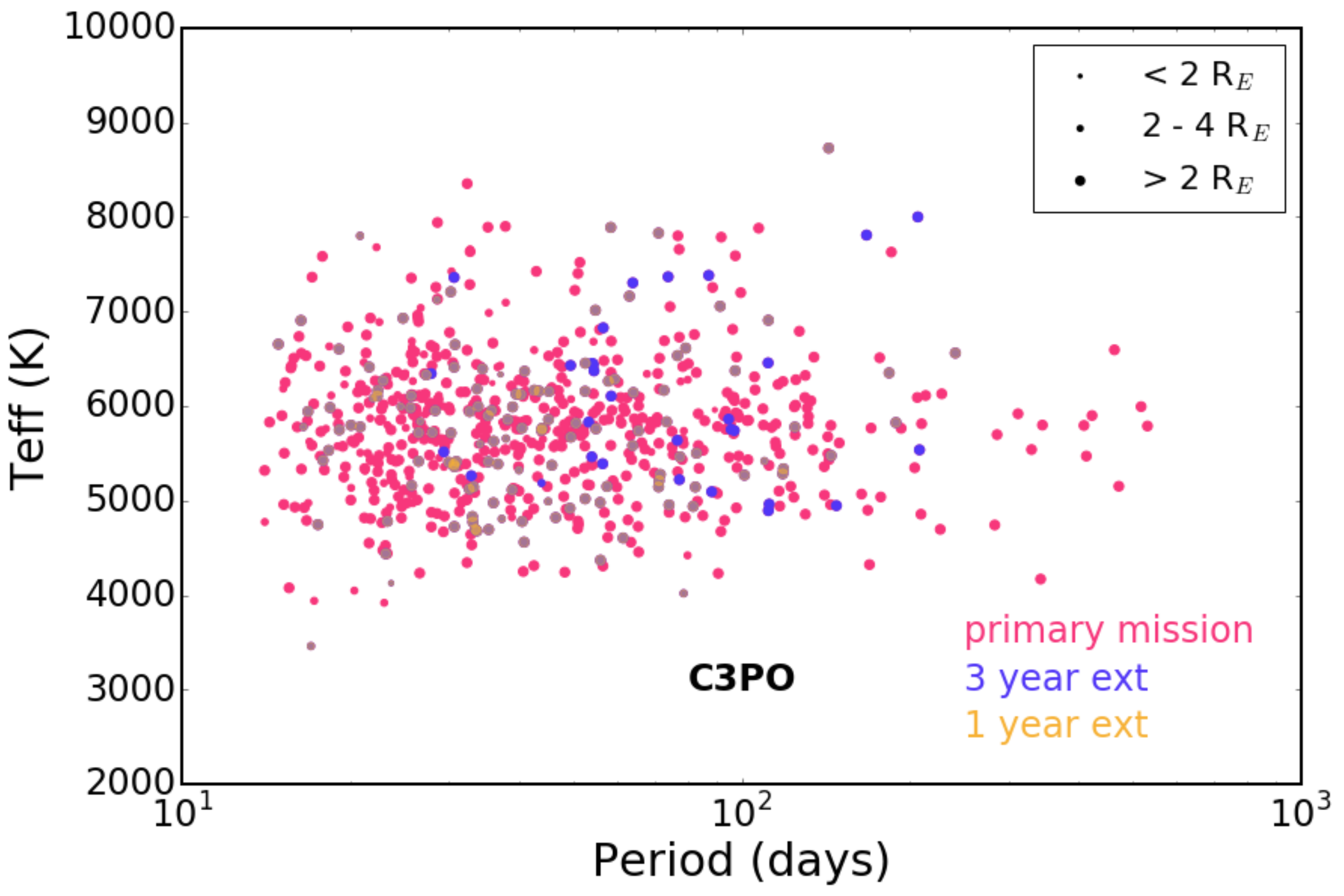}
    \caption{The distribution of FFI planets with a single primary mission transit as a function of \teff\ and orbital period. Pink points represent planets with a single transit discovered in the primary TESS mission, while orange and blue points mark systems where additional transits are detected in a 1-year and 3-year extended mission, respectively. The point size scales with planet radius. From top to bottom, we show the results for the {\bf Allsky}, {\bf Hemi}, and \asypole\ scenarios. \label{fig:Ext_Singles}}
\end{figure}

\begin{figure*}
    \centering
    \includegraphics[width=\textwidth]{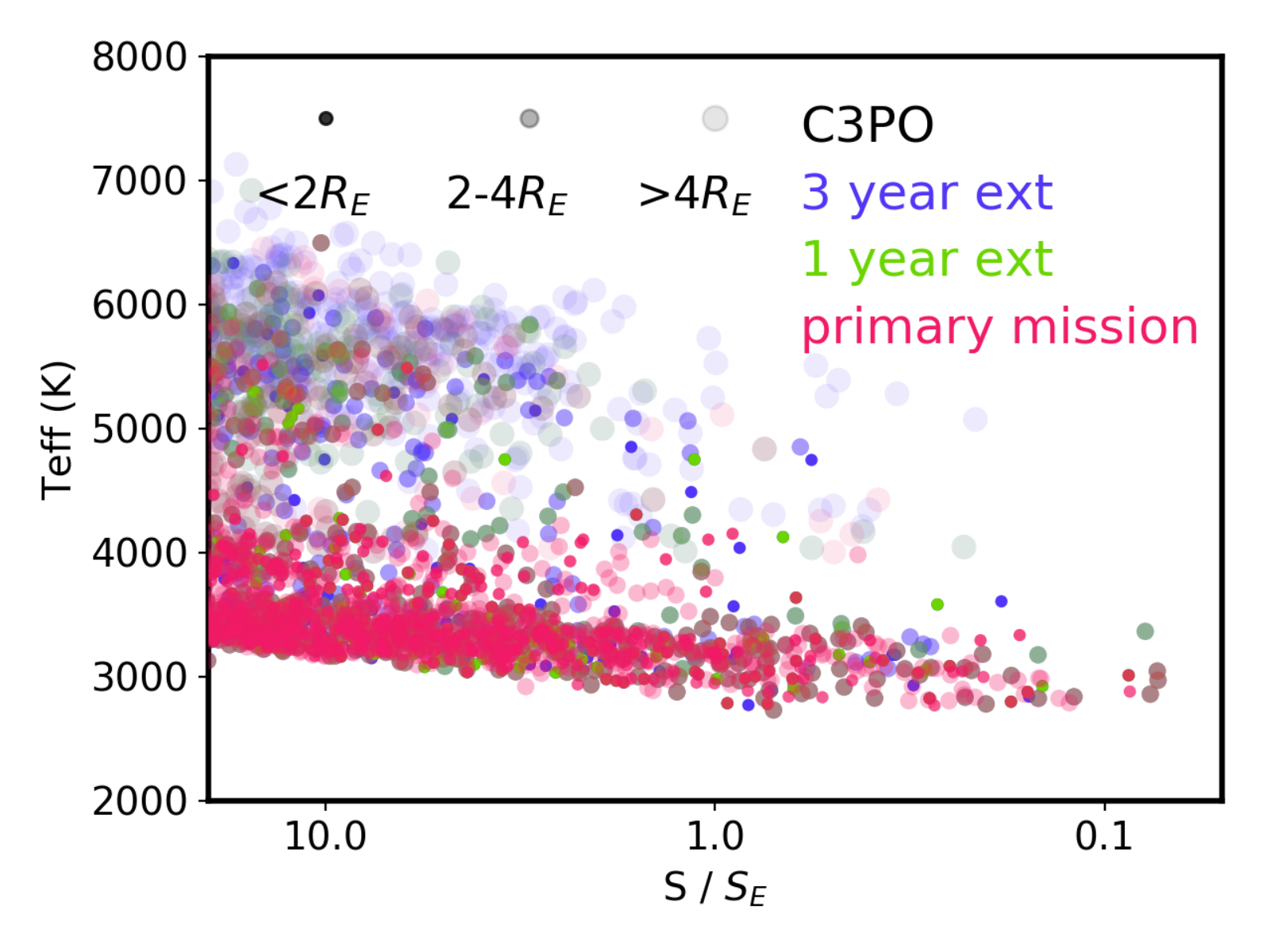}
    \caption{Relative irradiation of the planets $S/S_E$ versus the effective temperature of the host star for the "C3PO" extended mission. The figure zooms in near the region around $S/S_E=1$. The size of the point represents the size of the planets. The larger planets are more transparent.}
    \label{fig:Ext_Flux}
\end{figure*}

We expect a number of the single transits captured during the primary mission to be re-observed during an extended mission. However, we note that simply recovering one other transit (many orbital cycles after the first) is not sufficient to uniquely constrain the period (the period will only be constrained to a set of values such that there are an integer number of cycles between the two transits). For this reason, in order to confirm single-transit planets, it will be necessary to obtain non-TESS follow-up observations (e.g. radial velocity measurements) even for planets with re-observed transits.
For each of the three scenarios, Figure \ref{fig:Ext_Singles} shows in pink the distribution of FFI planets with a single primary mission transit; those recovered during a 1-year and 3-year extended mission are layered in orange and blue, respectively. The least successful scenario in terms of the recovery of new transits for the single transit planets observed in the primary mission is {\bf C3PO}. After one year, the number of single-transit planets with re-observed transits is comparable for the {\bf Hemi} and {\bf Allsky} scenarios. After three years, {\bf Allsky} becomes the most successful scenario thanks to its coverage of both hemispheres. Most of the planets with $P > 100$ days will not show new transits in any of the three extended mission scenarios studied here. The number of primary mission single-transit planets that remain un-observed after a three-year extended mission is 151, 396, and 530, for the {\bf Allsky}, {\bf Hemi}, and \asypole\ scenarios, respectively. The latter two numbers are dominated by planets from the un-observed hemisphere.

We note that the {\bf Hemi} and {\bf C3PO} scenarios are only investigated for the southern hemisphere because our knowledge of the pointing is better for the south at the current stage. Based on the estimation for the primary mission, we anticipate that an equivalent survey strategy for the northern hemisphere should produce similar results.
Selection of a hemisphere may be based on the availability of follow-up resources at the time of the extended mission.

\subsection{Caveats}
We caution the readers to be aware of the following caveats when interpreting the expected planet yields from this work.

(1) Target selection: We used a TP target list made prior to the TESS launch. In reality, the 2-minute cadence targets will be selected before the observation of each sector. The pointing plan for the northern hemisphere is more uncertain at this stage of the mission. We also do not include any GI targets (since they are not public) in the target selection process. 
Therefore, the targets determined to be on TESS CCDs in this work will not necessarily be observed. 

(2) Stellar parameters: 
Stars fainter than \tmag$\,= 12$ mag have less reliable stellar parameters. We did not cross check these stars with Gaia DR2 and some of them do not have an estimated flux contamination ratio. 
We also caution the readers regarding the expected number of planets around late M dwarfs ($T<3,100$ K). The planet yields around these stars are highly sensitive to the estimated stellar radii, which are not well calibrated. {\bf The radius estimation method we adopted introduces an intrinsic model dependent uncertainty $\sim 13.4\%$ on the stellar radius. Combined with the typical uncertainty of \teff (80 K) for the low mass stars in TIC, the overall uncertainty on the radius of low mass stars is approximately $15\%-20\%$.} 

(3) Planet injection:
The injected planet population mostly reflects our understanding of the exoplanet population in the {\em Kepler} field. The occurrence rates have larger uncertainties for some parts of parameter space such as: giant planets, small planets at long periods ($>$200 day), and planets around evolved stars. The occurrence rates we used in our simulation were not corrected for biases due to unresolved binaries in the Kepler survey \citep{Bouma:2018}, nor did we simulate the effect of binaries in our study. 
We assumed zero eccentricity for the orbit of every planet in our simulation. For planets with a noticeable amount of eccentricity, the transit duration is more likely to be shorter, but the transit probability will be larger. A fraction of giant planets at relatively long period have non-negligible eccentricities. Our treatment tends to underestimate this giant planet population.  

(4) Noise model: 
Our photometric noise model does not depend on stellar color. In reality, the PSF and QE of TESS have some color dependency, the amplitude of this effect will be measured in commissioning data. We also did not include potential degradation of data quality due to scattered light. \citet{Bouma:2017} estimated that the scattered light effect will lead to $\sim7\%$ reduction of the yield for planets by dropping the sectors suffering from Earth/Moon crossings completely, which is likely to be a upper limit. A more robust estimation on the effect of scattered light will also be estimated during commissioning.

\section{Discussion and summary}
\label{sec:discussion}

We simulated planet yields for the TESS 2-year primary mission, as well as three extended mission scenarios. Our simulations take advantage of realistic constraints based on a pre-launch optical model, mission planning profile, and current best knowledge of the stars (Gaia DR2) and planets ({\it Kepler} mission) TESS is going to observe.  

We report $\sim 10^4$ planets are likely to be discovered by the TESS primary mission, as well as an additional $\sim 2000$ planets for each year of the three extended mission scenarios we explored. We predict that in the primary mission, TESS will discover about 3500 planets of Neptune size or smaller, within which, $\sim$100 will have radii smaller than $1.25\,R_E$. Approximately 30 of the Earth-size planets will revolve around stars brighter than \tmag $\sim$ 10 mag. Our estimates show that the TESS primary mission, and its extended mission will greatly expand the sample of small planets that are temperately irradiated \footnote{The sample is defined by $R_p < 4\,R_E$, $0.32<S_p/S_E$<1.78}. To date, there are 24 planets satisfying such criteria in the exoplanet archive. At the end of a three year extended mission using the \asypole\ scenario, TESS is expect to find $\sim 250$ such planets, $\sim 30$ of which will revolve around F, G or K stars.    

We also find that an extended TESS mission will be beneficial for recovering additional transits of planets that show only one transit in the primary mission data. The three scenarios perform differently in this respect, with the {\bf allsky} and {\bf \asypole} scenarios recovering the largest and smallest number of single-transiting primary mission planets, respectively.

We highlight some of the science cases where TESS planets will enable a detailed investigation.

(1) Study the density diversity of small planets:
To date, there are fewer than 20 small planets ($R<4\,R_E$) with measured density uncertainty smaller than $20\%$. Typical small planets from TESS are expected to have radius uncertainties better than $5\%$ (given a combined constraint from Gaia and reconnaissance spectroscopy of the host star). A few hundred TESS small planets are expected to be hosted by stars with \tmag\ $<$ 10, potentially allowing accurate mass measurements (for slowly rotating and quiet host stars). \citet{Fulton:2017}, \citet{Fulton:2018}, and \citet{Berger:2018} have identified a dichotomy in the radius of small planets. Obtaining the densities of planets in the same radius range will greatly enlighten the understanding of the physical origin of such a dichotomy.

(2) Study planets around small and big stars: With the next generation of infrared high precision spectrographs, it is of special interest to identify close-in planets around M dwarfs to study their properties. TESS is expected to find an order of magnitude more planets around M dwarfs than the {\it Kepler} mission. TESS will also discover many planets around the brightest and most massive stars, which will provide intriguing new opportunities to understand extreme planetary atmospheres \citep{Shporer:2014, Beatty:2017, Gaudi:2017}.

(3) Detailed characterization of multiple planet systems: TESS will provide hundreds of new multiple transiting systems around bright stars. To date only a dozen multiple transiting planetary systems are known with host stars brighter than \tmag $=$ 12, and few have mass and other properties measured for all the planets around the same host star. Multiple-transiting planetary systems from TESS will enable follow-up observations to measure host star obliquities and individual planet densities, atmospheres, and orbital eccentricities, leading to a better understanding of planet formation and evolution.

\acknowledgments{
This work has made use of information from the TESS Mission. We thank the members of the TESS Science Team for their contributions to the mission. 
D. Dragomir acknowledges support provided  by NASA through Hubble Fellowship grant HST-HF2-51372.001-A awarded by the Space Telescope Science Institute, which is operated by the Association of Universities for Research in Astronomy, Inc., for NASA, under contract NAS5-26555. This work has made use of data from the European Space Agency (ESA)
mission {\it Gaia} (\url{https://www.cosmos.esa.int/gaia}), processed by
the {\it Gaia} Data Processing and Analysis Consortium (DPAC,
\url{https://www.cosmos.esa.int/web/gaia/dpac/consortium}). Funding
for the DPAC has been provided by national institutions, in particular
the institutions participating in the {\it Gaia} Multilateral Agreement.

}


\clearpage

\begin{turnpage}
\begin{table}
\caption{Simulated TESS planets detected in FFIs.\\
The table in its entirety is available in the online version of the journal, \\ only the first ten rows are given here for guidance of its content.
\label{tab:planets}}
\begin{tabular}{ccccccccccccccccccc}
\hline\hline
StarNo \ \ \ \ & \tmag  & R  & M & \teff & $\log g$ & R.A.  & Dec. & $f$ & $sector_s$ & P  & $R_p$  & b & $T_{\rm 14}$  & $N_p$ & $N_t$ & $\mu$ & $SNR_s$ & $s_d$ \\
& (mag) & ($R_\odot$) & ($M_\odot$) & (K) & (cgs) &  (deg) & (deg) & & & (day) & ($R_E$) & & (hr) &  &  & (rad) &  &  \\
\hline
1 & 10.14 & 1.28 & 1.3 & 5881 & 4.34 & 52.355408 & 10.900207 & 0.01 & 0000010000000 & 2.86 & 2.8 & 0.94 & 3.02 & 5 & 1.0 & 0.01 & 9.4 & 6 \\
\hline
2 & 9.84 & 0.88 & 1.01 & 5634 & 4.55 & 47.422302 & 10.483473 & 0.26 & 0000010000000 & 7.346 & 13.1 & 0.02 & 3.09 & 1 & 1.0 & 0.0 & 35.68 & 6 \\
\hline
3 & 7.5 & 1.32 & 1.2 & 6448 & 4.27 & 359.675725 & -38.225932 & 0.0 & 0011000000000 & 3.345 & 3.0 & 0.48 & 3.37 & 5 & 1.0 & 0.02 & 18.53 & 3 \\
\hline
4 & 11.33 & 1.73 & 0.98 & 5462 & 3.95 & 2.425111 & -36.95304 & 0.0 & 0001000000000 & 1.571 & 11.9 & 0.67 & 3.68 & 1 & 1.0 & 0.0 & 63.3 & 4 \\
\hline
5 & 5.49 & 4.12 & 1.32 & 5605 & 3.33 & 2.337712 & -27.988037 & 0.0 & 0001000000000 & 2.177 & 3.9 & 0.32 & 8.82 & 5 & 2.0 & 0.02 & 7.21 & 4 \\
\hline
6 & 5.49 & 4.12 & 1.32 & 5605 & 3.33 & 2.337712 & -27.988037 & 0.0 & 0001000000000 & 6.684 & 3.3 & 0.68 & 12.82 & 5 & 2.0 & 0.01 & 5.08 & 4 \\
\hline
7 & 10.3 & 1.29 & 1.01 & 5916 & 4.22 & 351.06367 & -31.23967 & 0.0 & 0010000000000 & 7.968 & 3.5 & 0.39 & 4.67 & 5 & 1.0 & 0.02 & 6.43 & 3 \\
\hline
8 & 9.64 & 2.5 & 0.85 & 5159 & 3.57 & 357.638148 & -25.936884 & 0.0 & 0001000000000 & 1.753 & 8.0 & 0.41 & 5.78 & 1 & 1.0 & 0.0 & 30.06 & 4 \\
\hline
9 & 9.75 & 2.12 & 1.24 & 6242 & 3.87 & 357.898245 & -23.745236 & 0.0 & 0001000000000 & 4.814 & 3.8 & 0.67 & 6.05 & 3 & 1.0 & 0.02 & 7.04 & 4 \\
\hline\hline
\end{tabular}
\end{table}

\end{turnpage}

{}

\end{document}